\documentclass[5p]{elsarticle}

\usepackage{subcaption}
\usepackage{pgfplots}
\usepackage{color}
\usepackage{amssymb}
\usepackage{amsmath, cases}
\usepackage{booktabs}
\usepackage{lineno,hyperref}

\makeatletter
\def\ps@pprintTitle{%
	\let\@oddhead\@empty
	\let\@evenhead\@empty
	\def\@oddfoot{}%
	\let\@evenfoot\@oddfoot}
\makeatother

\newtheorem{definition}{Definition}

\begin{document}
	
	\begin{frontmatter}
		
		\title{Fast extraction of the backbone of projected bipartite networks to aid community detection} 
		
		\author[mymainaddress]{J. Liebig} %\corref{mycorrespondingauthor}}
		\author[mymainaddress]{A. Rao}
		\address[mymainaddress]{School of Mathematical and Geospatial Sciences, RMIT University, Melbourne 3001, Australia}
		
		\begin{abstract}
			This paper introduces a computationally inexpensive method of extracting the backbone of one-mode networks projected from bipartite networks. We show that the edge weights in the one-mode projections are distributed according to a Poisson binomial distribution and that finding the expected weight distribution of a one-mode network projected from a random bipartite network only requires knowledge of the bipartite degree distributions. Being able to extract the backbone of a projection is highly beneficial in filtering out redundant information in large complex networks and narrowing down the information in the one-mode projection to the most relevant. We demonstrate that the backbone of a one-mode projection aids in the detection of communities.
		\end{abstract}
		
		
	\end{frontmatter}
	
% % % % % % % % % % % % % % % % % % % % % % % % % % % % % % % % % % % % % % % % 
\section{Introduction}\label{sec:intro}
% % % % % % % % % % % % % % % % % % % % % % % % % % % % % % % % % % % % % % % %
Scientist are often faced with an overwhelmingly large amount of data. Being able to reduce this to the most relevant information greatly simplifies analyses. In network science, a way of reducing networked data is discarding redundant or insignificant edges, resulting in a network that is commonly referred to as the backbone. The backbone of a network $\mathcal{G}(V,E)$, with node set $V$ and edge set $E$, is defined as the subgraph $\mathcal{G}'(V,E')$ of $\mathcal{G}$, such that the edge set $E'$ of the backbone $\mathcal{G}'$ contains only the most significant edges in $E$. Despite this straightforward definition, many challenges lie in the identification of significant edges.

Many different approaches to determining the most significant edges of a given network have been proposed, all pursuing the same goal of retaining only the most relevant information~\cite{Glattfelder2009, Slater2009, Scutari2013, Zhang2014}. Most methods focus on weighted one-mode networks, identifying statistically significant edges by comparison to, for example, a null-model \cite{Serrano2009} or edge weight distributions~\cite{Foti2011}. Although the aim of backbone extraction is reduction of data in order to simplify analysis, it is vital that the backbone still contains important and relevant information about the network. Hence, it is crucial that topological structures are preserved in the backbone~\cite{Zhang2013b}.  

Comparable approaches have been suggested for one-mode projections~\cite{Neal2013, Neal2014}. A one-mode projection is the projection of a bipartite network onto a one-mode network. Bipartite networks consist of two disjoint sets of nodes $U$ and $V$, where edges only connect nodes belonging to different sets. A bipartite network is usually represented by its biadjacency matrix $B$, where the element $b_{ij}=1$ if node $u_i\in U$ is connected to node $v_j\in V$ and 0 otherwise. The one-mode projection of this bipartite network is constructed by dropping one node set and connecting two nodes of the remaining set if they share at least one neighbour in the bipartite network. Edges connecting the two nodes in the one-mode projection are associated with a weight, namely the number of previously shared neighbours.

Bipartite networks are far less studied than ordinary one-mode networks and hence many network measures are not applicable to bipartite structures. Although some measures have been redefined to suit the analysis of bipartite networks~\cite{Borgatti1997,Liebig2014,Liebig2016}, the study of their projections is preferred in most cases~\cite{Newman2001a}.

Being able to infer the backbone of one-mode projections is especially important as the process of projection results inevitably in a dense and noisy one-mode network~\cite{Neal2013}. The extracted backbone could also be translated into a binary network, as it is often preferable to work with an unweighted version of the network~\cite{Latapy2008}.

Neal~\cite{Neal2014} introduced a stochastic degree sequence model to determine the significance of edges in weighted projections. In contrast to previous approaches,~\cite{Neal2014} takes the degrees of primary and secondary nodes in the original bipartite network into consideration. Determining the significant edges requires the generation of a reasonably large set of random bipartite networks and their one-mode projections. The random projections serve to determine the expected weight distribution of every edge in the network. Comparing the observed edge weights of the projections to their corresponding expected weight distribution allows the identification of significant edges. 

The one-mode projection of a bipartite network is obtained by multiplying its biadjacency matrix $B$ with its transpose. Hence, the process of projecting a bipartite network is computationally expensive, running in $\mathcal{O}(|U|^2|V|)$ time and as~\cite{Neal2014} opines, a method to directly calculate edge weight distributions would be highly beneficial. In this paper we do exactly this, showing that the edge weights in most random one-mode projections are distributed according to a Poisson binomial distribution. We develop a fast method of extracting the backbone of a one-mode projection that does not rely on the generation of random networks and their projections and hence significantly reduces computation time. We corroborate the accuracy of our method by comparing the weights thus obtained to the real weight distribution of the projections of several types of randomly generated bipartite networks.

Further, we demonstrate, by means of two real world networks, that backbone extraction aids in the detection of communities. A community is loosely defined to be a subgraph of a network with a relatively higher number of inner connections compared to the number of edges linking to nodes outside the subgraph. We show that communities are more pronounced in the backbone than in the projection. This is a very interesting observation, since many real world networks exhibit community structure.

% % % % % % % % % % % % % % % % % % % % % % % % % % % % % % % % % % % % % % % % 
\section{The weight distribution of one-mode projections}\label{sec:weightDistribution}
% % % % % % % % % % % % % % % % % % % % % % % % % % % % % % % % % % % % % % % %
In the following, we show that the weight distribution of most projected random bipartite networks follows a Poisson binomial distribution. We give some remarks on exceptions at the end of this section.

% % % % % % % % % % % % % % % % % % % % % % % % % % % % % % % % % % % % % % % %
\subsection{The Poisson Binomial Distribution}
We begin with the necessary definitions and notation.

\begin{definition}~\cite{Forbes2011}
	A Bernoulli trial is a random variable $X$ with two possible outcomes, success or failure, and is associated with a success probability $p$.
\end{definition}

The probability of obtaining $n$ successes in $N$ independent Bernoulli trials, where each trial has success probability $p$, is given by the binomial distribution~\cite{Forbes2011}: 

\begin{equation}\label{eqn:binomial}
	P(X_1+\dots +X_N=n) = \left(\begin{array}{l}
		N\\
		n
	\end{array}\right)p^n(1-p)^{N-n}.
\end{equation} 

If the $N$ trials have varying probabilities $p_i$, where $i=1,\dots, N$, the sum of the independent, non-identically distributed random variables $X_1,\dots, X_N$ is given by the Poisson binomial distribution~\cite{Wang1993}. 

Let $\mathcal{S}_n$ be the set of all combinations of $n$ distinct integers chosen from $\{1,\dots, N\}$ and let $S_1, \dots, S_{|\mathcal{S}_n|}$ be the elements of $\mathcal{S}_n$. Let $s$ denote an element of the subset $S_j$, where $1\leq j\leq |\mathcal{S}_n|$ and let $\bar{S}_j$ denote the complement of $S_j$ with respect to $\{1,\dots, N\}$. Then the probability density function of the Poisson binomial random variable $Z_X=\sum\limits_{i=1}^{N}X_i$ is given by 

\begin{equation}
	P(Z_X = n) =  \sum\limits_{j=1}^{|\mathcal{S}_n|}\prod\limits_{s\in S_j}p_s \prod\limits_{\bar{s}\in\bar{S}_j}(1-p_{\bar{s}}).
\end{equation}

% % % % % % % % % % % % % % % % % % % % % % % % % % % % % % % % % % % % % % % %
\subsection{Approximation of the weight distribution}
We now look at the use of the Poisson binomial distribution to approximate the weight distribution. 

\begin{definition}~\cite[Definition~1.4]{Lando2003}\label{def:generating}
	Let $a_0, a_1, a_2, \dots$ be an arbitrary (infinite) sequence of numbers. The generating function for this sequence is the expression $\sum\limits_{n=0}^{\infty}a_nx^n$.
\end{definition}

Let $\mathcal{B}$ be a bipartite network with the two disjoint node sets $U$ and $V$, where $U$ is called the primary set and $V$ the secondary set. Let $p_j$ denote the probability that a node $u\in U$ has degree $j$ and $q_k$ denote the probability that a node $v\in V$ has degree $k$.

By Definition \ref{def:generating}, $f(x) = \sum_{j=0}^{\infty}p_jx^j$ is the probability generating function of the primary node degrees and $g(x)= \sum_{k=0}^{\infty}q_kx^k$ is the probability generating function of the secondary node degrees. It is important to note that 

\begin{equation}
	f(1) = \sum\limits_{j=0}^{\infty}p_j = 1
\end{equation} 

and 

\begin{equation}
	\left.\left(x\frac{d}{dx}\right)^nf(x)\right|_{x=1} = \sum\limits_{j=0}^{\infty}j^np_j=\langle j^n\rangle,
\end{equation}

where $\langle j\rangle$ denotes the average node degree of the primary node set~\cite[Chapter~13]{Newman2010}. Note that $\langle j^n \rangle \neq \langle j \rangle^n$.

The probability of an edge connecting a primary node to a secondary node is then determined by dividing the product of their degrees by the number of edges in the network. The number of edges $m$ in a bipartite network is given by   

\begin{equation}
	m = |U|\langle j\rangle = |V|\langle k \rangle.
\end{equation}

If $\pi_{uu'v}$ denotes the probability that two primary nodes $u$ and $u'$ are connected to a secondary node $v$, then given that $deg(u)=j_u$, $deg(u')=j_{u'}$ and $deg(v)=k_v$,

\begin{eqnarray}
	\pi_{uu'v} &=& \frac{j_uj_{u'}k_v(k_v-1)}{m(m-1)}\nonumber \\
	&=& \frac{j_uj_{u'}k_v(k_v-1)}{|U|^2\langle j \rangle^2-|U|\langle j\rangle}.
\end{eqnarray}  

Averaging $j_u$ and $j_{u'}$ over the probabilities of an edge attaching to any node of degree $j_u$ and $j_{u'}$ respectively, results in the probability $\pi_v$ that any two primary nodes are connected to a particular secondary node $v$ of degree $k_v$. Note that $p_j$ is the fraction of nodes with degree $j$ in the primary node set. Thus, the number of primary nodes with degree $j$ is equal to $|U|p_j$ and the probability that an edge originating at node $v$ connects to a primary node of degree $j$ is $|U|jp_j/m = jp_j/\langle j \rangle$~\cite[Chapter~13]{Newman2010}. Hence,

\begin{eqnarray}\label{eqn:probV}
	\pi_v &=& \sum\limits_{j_u, j_{u'}} \frac{j_u^2p_{j_u}j_{u'}^2p_{j_{u'}}k_v(k_v-1)}{(|U|^2\langle j \rangle^2-|U|\langle j\rangle)\langle j\rangle^2}\nonumber\\
	&=& \frac{k_v(k_v-1)}{|U|^2\langle j \rangle^4-|U|\langle j\rangle^3} \left[\sum_{j=0}^{\infty}j^2p_j\right]^2\nonumber\\
	&=& \frac{\langle j^2 \rangle^2 k_v(k_v-1)}{|U|^2\langle j \rangle^4-|U|\langle j\rangle^3}. 
\end{eqnarray}

$\pi_v$ is the probability of the Bernoulli random variable $X_v$ indicating the existence of a connection between two primary nodes via a particular secondary node $v$.   

It follows that the probability of a randomly chosen edge in the one-mode projection having weight $\omega$ is given by $P\left(\Omega_X=\sum_{v=1}^{|V|}X_v=\omega\right) = \sum_{j=1}^{|\mathcal{S}_\omega|} \prod_{s\in S_j}\pi_s \prod_{\bar{s}\in\bar{S}_j}(1-\pi_{\bar{s}})$, where $\mathcal{S}_\omega$ is the set of all combinations of $\omega$ integers chosen from $\{1,\dots, |V|\}$.

Since $P(\Omega_X=\omega)$ is hard to compute, we use the Poisson approximation instead: 

\begin{equation}\label{eqn:poisApprox}
	P(\Omega_X=\omega) \approx \frac{\mu^\omega e^{-\mu}}{\omega!},
\end{equation} 

where $\mu =\sum_{v=1}^{|V|} \pi_v$.

The error of the Poisson approximation to $P(\Omega_X=\omega)$ is $\epsilon_\omega < 2\sum_{i=1}^{|V|}\pi_{v}^2$ and is small if the number of expected successes is small~\cite{LeCam1960}. Since most real world networks are sparse, the Poisson approximation estimates the probability of weight $\omega$ very well.

Notice that $\mu$ is also easily calculated: 

\begin{eqnarray}\label{eqn:mu}	
	\mu &=& \sum\limits_{v=1}^{|V|} \pi_{v}\nonumber\\
	&=& \frac{\langle j^2 \rangle^2}{|U|^2\langle j \rangle^4-|U|\langle j\rangle^3} \sum\limits_{v=0}^{|V|} k_{v}(k_{v}-1) \nonumber\\
	&=& \frac{|V|\langle j^2 \rangle^2\left(\langle k^2\rangle -\langle k \rangle\right)}{|U|^2\langle j \rangle^4-|U|\langle j\rangle^3}.
\end{eqnarray}  

% % % % % % % % % % % % % % % % % % % % % % % % % % % % % % % % % % % % % % % %
\subsection{Determining probabilities of individual connections}
When extracting the backbone of a network, one is interested in the probability of observing a connection with a certain weight between two nodes $u$ and $u'$ in the projection, denoted by $P_{uu'}(\Omega_X=\omega)$.

If $\pi_{uu'v}$ is small for every $v=1,\dots, |V|$, the Poisson approximation may be used, with 

\begin{eqnarray}\label{eqn:muInd}
	\mu &=& \sum\limits_{v=1}^{|V|}\pi_{uu'v}\nonumber\\
	&=& \sum\limits_{v=1}^{|V|} \frac{j_uj_{u'}k_{v}(k_{v}-1)}{|U|^2\langle j \rangle^2 - |U|\langle j\rangle}\nonumber\\
	&=& \frac{|V|j_uj_{u'}\left(\langle k^2 \rangle - \langle k\rangle\right)}{|U|^2\langle j \rangle^2-|U|\langle j\rangle}.
\end{eqnarray}  

In bipartite networks where some nodes have a very high degree, it is often found that the probability of a connection between two individual nodes is very high, resulting in large approximation errors. In such situations, instead of the Poisson approximation, we use the normal approximation:

\begin{equation}\label{eqn:normalApprox}
	P_{uu'}(\Omega=\omega) \approx \frac{1}{\sigma\sqrt{2\pi}}e^{-(\omega-\mu)^2/(2\sigma^2)},
\end{equation}

where $\mu$ is given by Eq.~(\ref{eqn:muInd}) and

\begin{eqnarray}\label{eqn:sigmaInd}
	\sigma &=& \left[\sum\limits_{v=1}^{|V|}\pi_{uu'v}(1-\pi_{uu'v})\right]^{1/2}\nonumber\\
	&=& \left[\sum\limits_{v=1}^{|V|}\pi_{uu'v} - \sum\limits_{v=1}^{|V|}\pi_{uu'v}^2\right]^{1/2}\nonumber\\
	&=& \left[\mu - \sum\limits_{v=1}^{|V|} \left(\frac{j_uj_{u'}k_v(k_v-1)}{|U|^2\langle j \rangle^2-|U|\langle j\rangle}\right)^2\right]^{1/2}\nonumber\\
	&=& \left[\mu - \frac{|V|j^2_uj^2_{u'}\left(\langle k^4\rangle - 2\langle k^3 \rangle + \langle k^2 \rangle\right)}{|U|^4\langle j\rangle^4-2|U|^3\langle j\rangle^3+|U|^2\langle j\rangle^2}\right]^{1/2}.
\end{eqnarray}

\begin{figure}[b]
	\includegraphics[width=80mm]{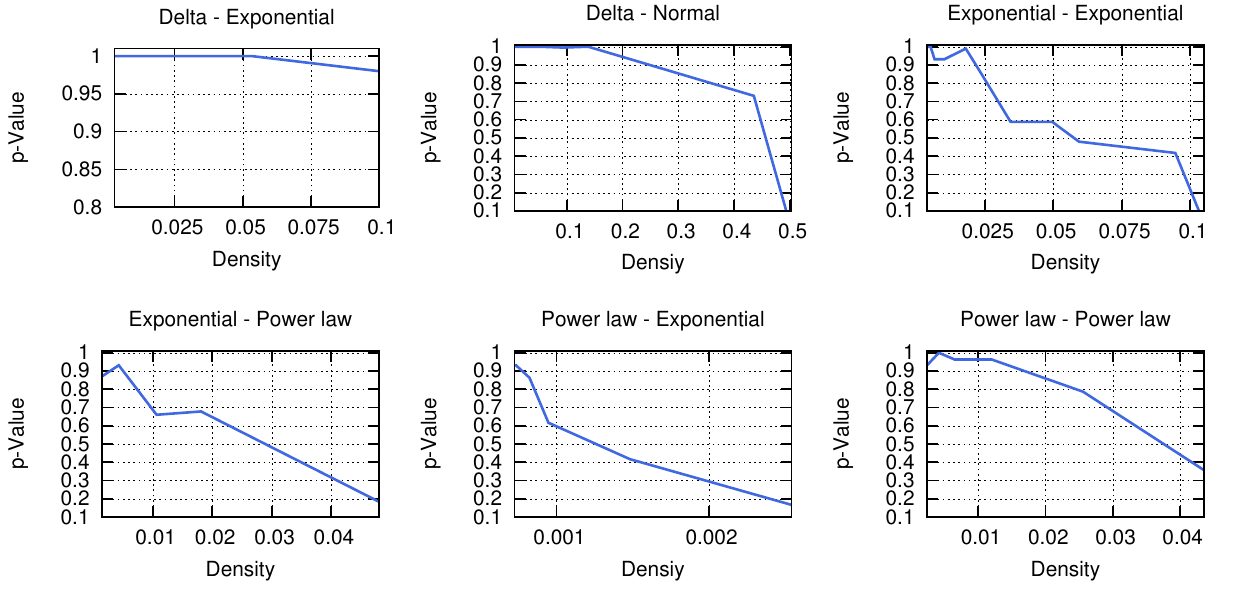}
	\caption{Results of the KS tests for six of the 25 tested cases.}\label{im:ks}
\end{figure}

In order to demonstrate the accuracy of our approximation, we considered bipartite networks from all 25 possible permutations of the following degree distributions: Delta function, uniform, normal, exponential and power law distributions. For each permutation, we generated 100 random bipartite networks of the same size and projected each onto a one-mode network to determine their average weight distribution. We used the Kolmogorov-Smirnov (KS) test to compare the observed average weight distribution with the approximation. In addition, we tested our approximation for robustness, by changing the distribution parameters. The results of the KS tests suggest that the approximation estimates the expected weight distribution of a projected bipartite network extremely well and is robust against variation of the distribution parameters. As expected, our approximation performs poorly only for very dense networks, with p-values falling below 0.5. As most real world networks are extremely sparse such cases are rarely observed~\cite{Konect2014, Newman2010}. Figure~\ref{im:ks} shows the results of the KS tests for six commonly observed pairs of degree distributions.

In the extraction of the backbone of a projected bipartite network, we are interested in the probabilities of connections between individual nodes rather than the complete weight distribution. The weight probability distribution of an edge between two individual nodes $u$ and $u'$ is given by Eq.~(\ref{eqn:normalApprox}). Since  $P_{uu'}(\Omega=\omega)$ has to be computed for every edge in the network, the computation time amounts to $\mathcal{O}(|U|^2)$. The greatest advantage of our method, however, is that it is not necessary to generate any random networks, saving the time required to generate and then project hundreds or thousands of networks. Since every single projection runs in $\mathcal{O}(|U|^2|V|)$ time our approach greatly simplifies and speeds up the process of extracting the backbone of a one-mode projection. 

% % % % % % % % % % % % % % % % % % % % % % % % % % % % % % % % % % % % % % % %
\section{Applications}\label{sec:examples}
% % % % % % % % % % % % % % % % % % % % % % % % % % % % % % % % % % % % % % % %

Next, we demonstrate our approach by extracting the backbone of two projected networks, the 108\textsuperscript{th} U.S. Senate network~\cite{Fowler2006, Fowler2006a} and the MovieLens Tag Genome network~\cite{Vig2012}. The first network has previously been analysed, revealing that the backbone of the projected bipartite network contains two communities of democrats and republicans respectively~\cite{Neal2014}. Our work compares the performance of community detection algorithms using the one-mode projection and the backbone. We show that a higher modularity is achieved if the backbone is used as an input. 

\begin{figure}[h]
	\centering
	\includegraphics[width=80mm]{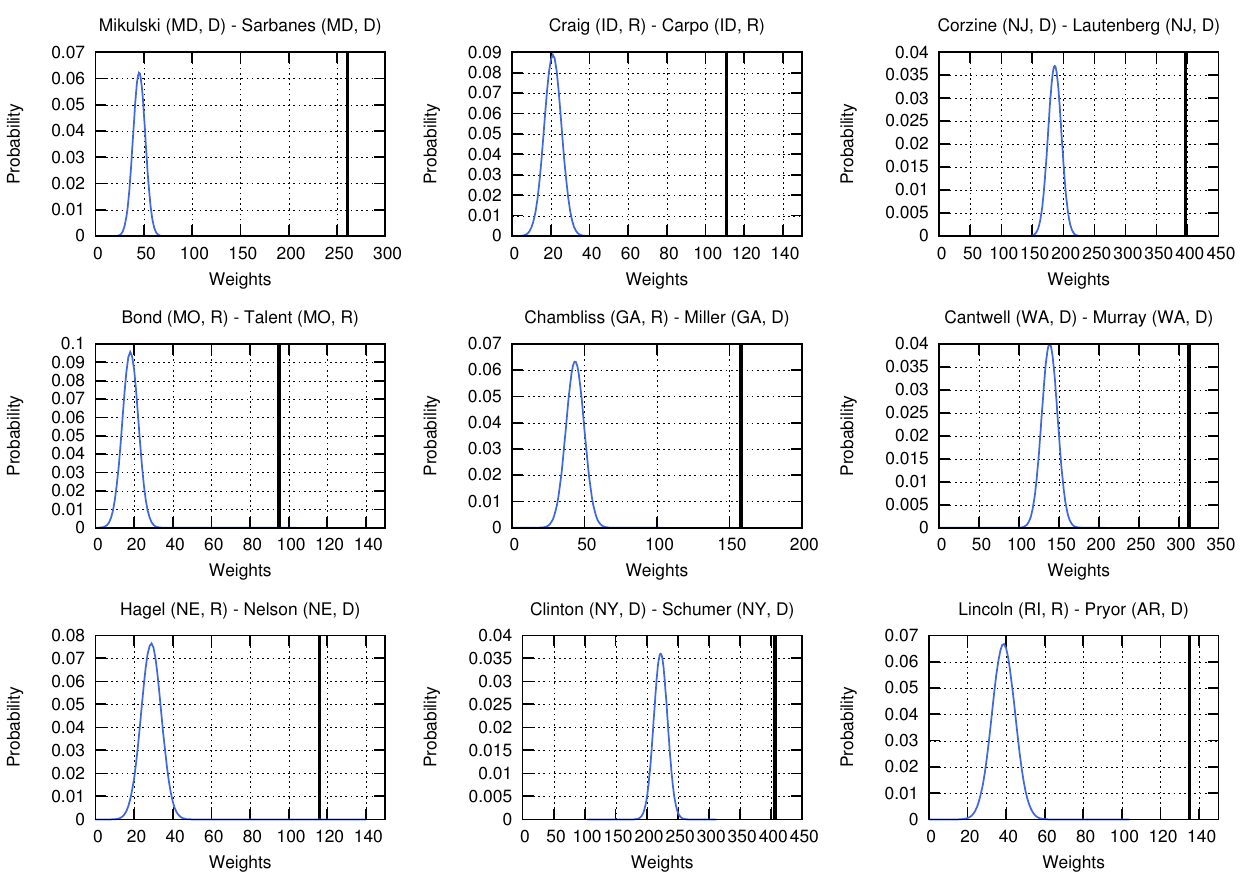}
	\caption{The weight probability distributions of the nine most significant edges in the senator-senator projection, where the observed weight is greater than expected. The blue curves show the approximated probability distributions, the black vertical bars mark the observed weight in the weighted one-mode projection of the 108\textsuperscript{th} U.S. Senate network.}\label{im:IndSignificant}
\end{figure}

There has been much interest in the detection of network communities and a large body of literature exists on algorithms detecting them. We choose to use the approach suggested in~\cite{Newman2006}, using eigenvectors, implemented in the R programming language~\cite{Csardi2006}. This eigenvector community detection algorithm aims to divide the input network into groups such that the modularity function $Q$, obtained by subtracting the number of expected edges within communities from the observed number of connections within communities, is maximised. Rewriting $Q$ in terms of matrices allows one to view the optimisation problem as a spectral problem, reducing its complexity, as community structure is often encoded in the first few eigenvectors of the modularity matrix. For more details see~\cite{Newman2006}.

% % % % % % % % % % % % % % % % % % % % % % % % % % % % % % % % % % % % % % % % 
\subsection{108\textsuperscript{th} U.S. Senate Data}
The U.S. Senate together with the House of Representatives builds the U.S. Congress. In every state two senators are voted into the senate and may introduce a piece of legislation, called bill. Bills can be co-sponsored by other members of the senate. The U.S. Senate data set may be represented as a bipartite network with 100 primary nodes, the senators, and 7,804 secondary nodes, the bills. An edge indicates that a senator has sponsored or co-sponsored a bill. This data set is publicly available and can be downloaded from http://jhfowler.ucsd.edu/cosponsorship.htm.  

\begin{figure}[h]
	\centering
	\includegraphics[width=80mm]{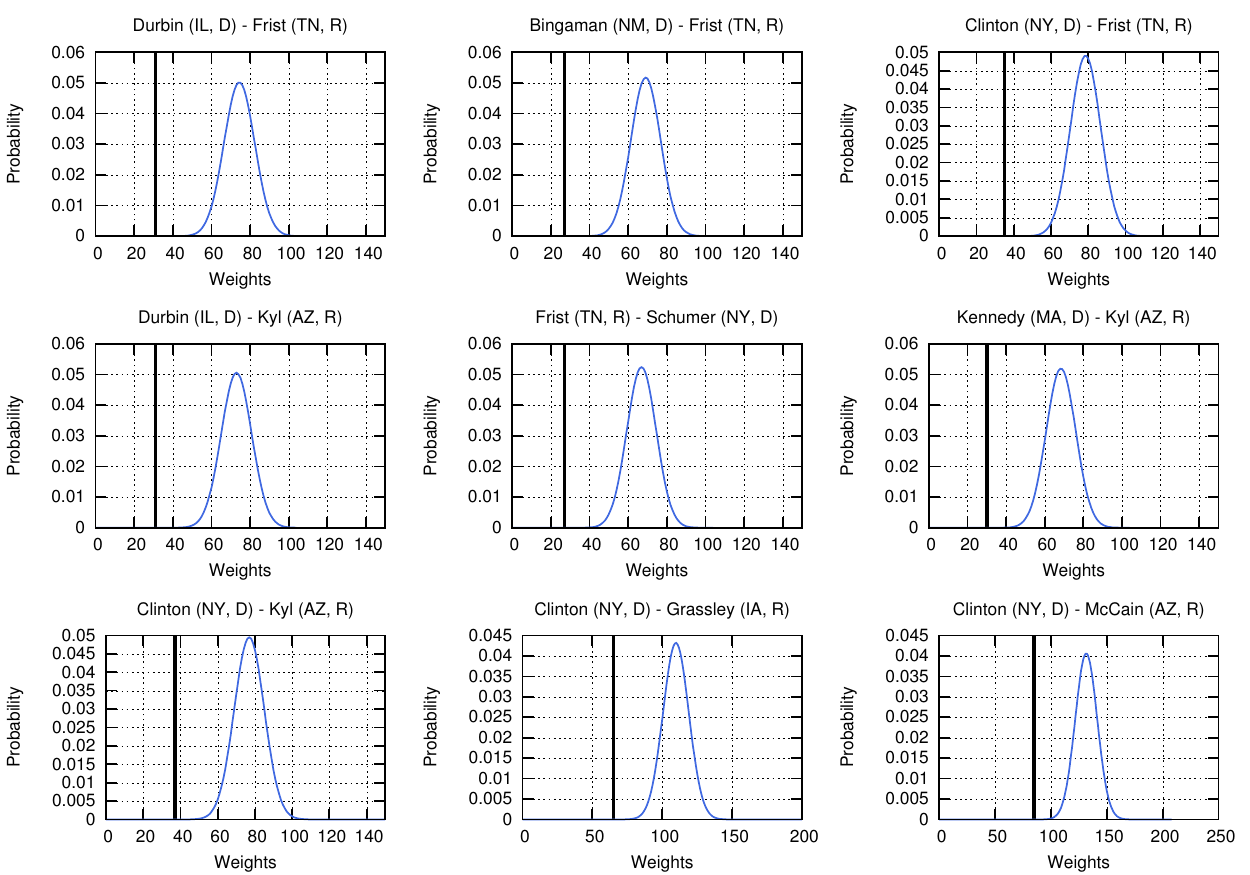}
	\caption{The weight probability distributions of nine edges in the senator-senator projection, where the observed weight is smaller than expected (not included in the backbone). These edges represent political antagonisms. The blue curves show the approximated probability distributions, the black vertical bars mark the observed weight in the weighted one-mode projection of the 108\textsuperscript{th} U.S. Senate network.}\label{im:IndInsignificant}
\end{figure}

In order to extract the backbone of the senator-senator network, we need to find the weight probability distribution for every edge between all possible pairs of senators, 4,950 in total. If the observed edge weight between two senators is significantly larger than expected, the unweighted edge is included in the backbone. Here, an edge is included in the backbone if its weight is greater than the mean plus three standard deviations of the approximated distribution. The threshold of three standard deviations corresponds to a 95\% confidence interval. This threshold is arbitrary and may be chosen differently. We found that lower thresholds attained lower modularities, whereas higher thresholds removed a larger number of edges, leading to the risk of deleting connections that are key to the network's topology. Although higher thresholds achieved higher modularities, the network simultaneously became more fragmented. For instance, choosing an extensive threshold of ten standard deviations resulted in a modularity of 0.57 and 27 groups of senators many of whom were isolated nodes. Albeit that the threshold is arbitrary, it has to be carefully chosen such that a high modularity is achieved while not fragmenting the network to the point of loss of important topological features.

Plots of the weight distributions of some of the most significant edges are shown in Figs.~\ref{im:IndSignificant} and~\ref{im:IndInsignificant}. 

\begin{figure}[h]
	\centering
	\includegraphics[width=80mm]{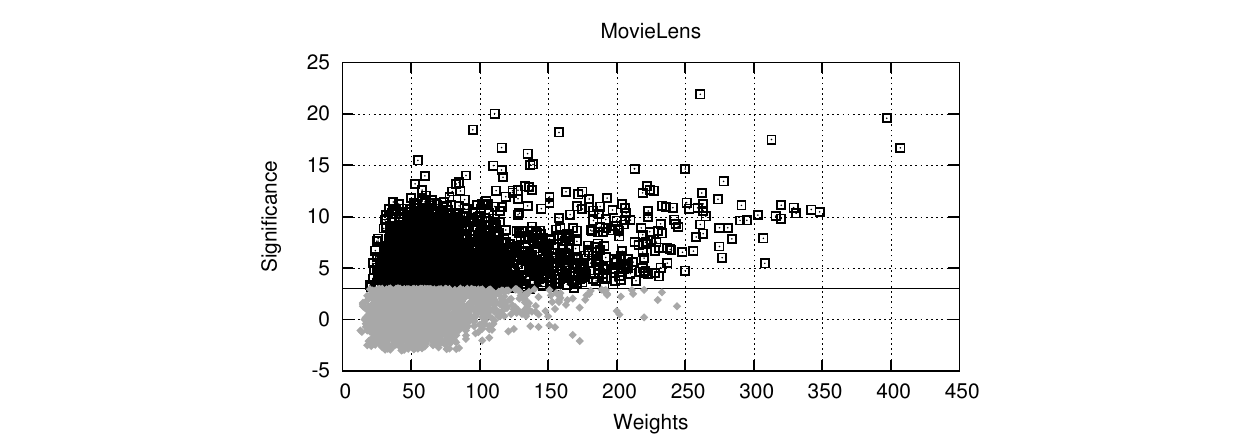}
	\caption{Edge significances in the U.S. Senate network plotted against their corresponding edge weights show no correlation between the two. Significant edges are represented by black squares.}\label{im:correlation}
\end{figure}

\begin{figure}[h]
	\centering
	\includegraphics[width=80mm]{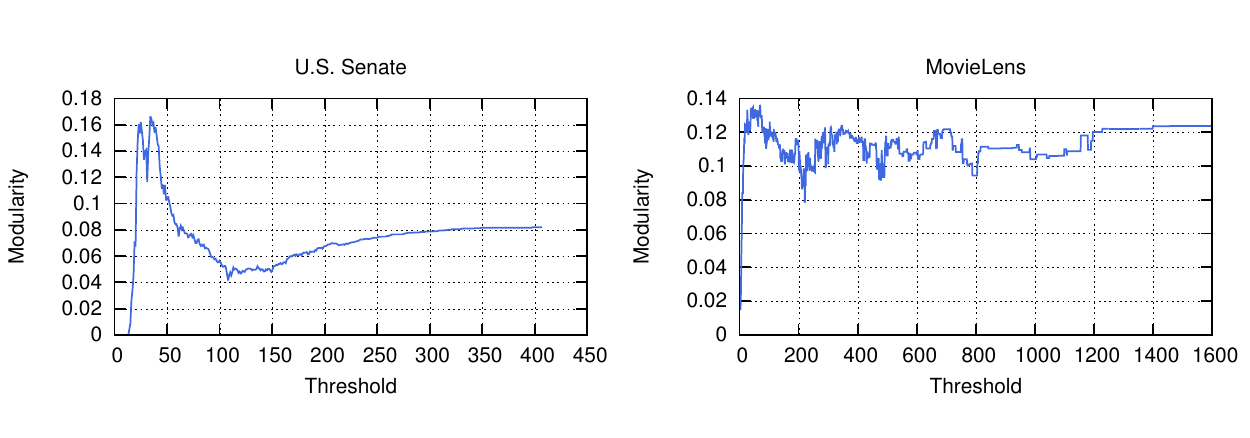}
	\caption{Extracting the backbone of the U.S Senate and the MovieLens networks by removing edges with weights under a certain threshold demonstrates that the increase in modularity achieved by our method (see below) is not trivial.}\label{im:modularity}
\end{figure}

Determining edge significance by comparing each edge weight individually to its expected distribution ensures that edges with high weights are not chosen over edges with low weights, for inclusion in the backbone~\cite{Neal2014}. Figure~\ref{im:correlation} illustrates that edge weights and edge significance are, with a correlation coefficient of 0.368, weakly correlated. Black squares indicate edges that were identified as being significant connections. Edge significance is calculated by subtracting the mean of the individual edge distribution from the observed weight and dividing by the distribution's standard deviation. It follows that the significance of edges cannot be trivially determined by removing all edges with a weight below a certain threshold. Consider for instance two senators with low degrees and assume they cosponsor the maximum possible number of bills, given by the minimum of the two degrees. Despite the edge connecting the two senators in the one-mode projection having a low weight, their relationship would be considered significant. Trivial edge removal may therefore lead to deletion of connections within communities and can never achieve high modularities, see Fig.~\ref{im:modularity}.

\begin{figure}[h]
	\centering
	\includegraphics[width=80mm]{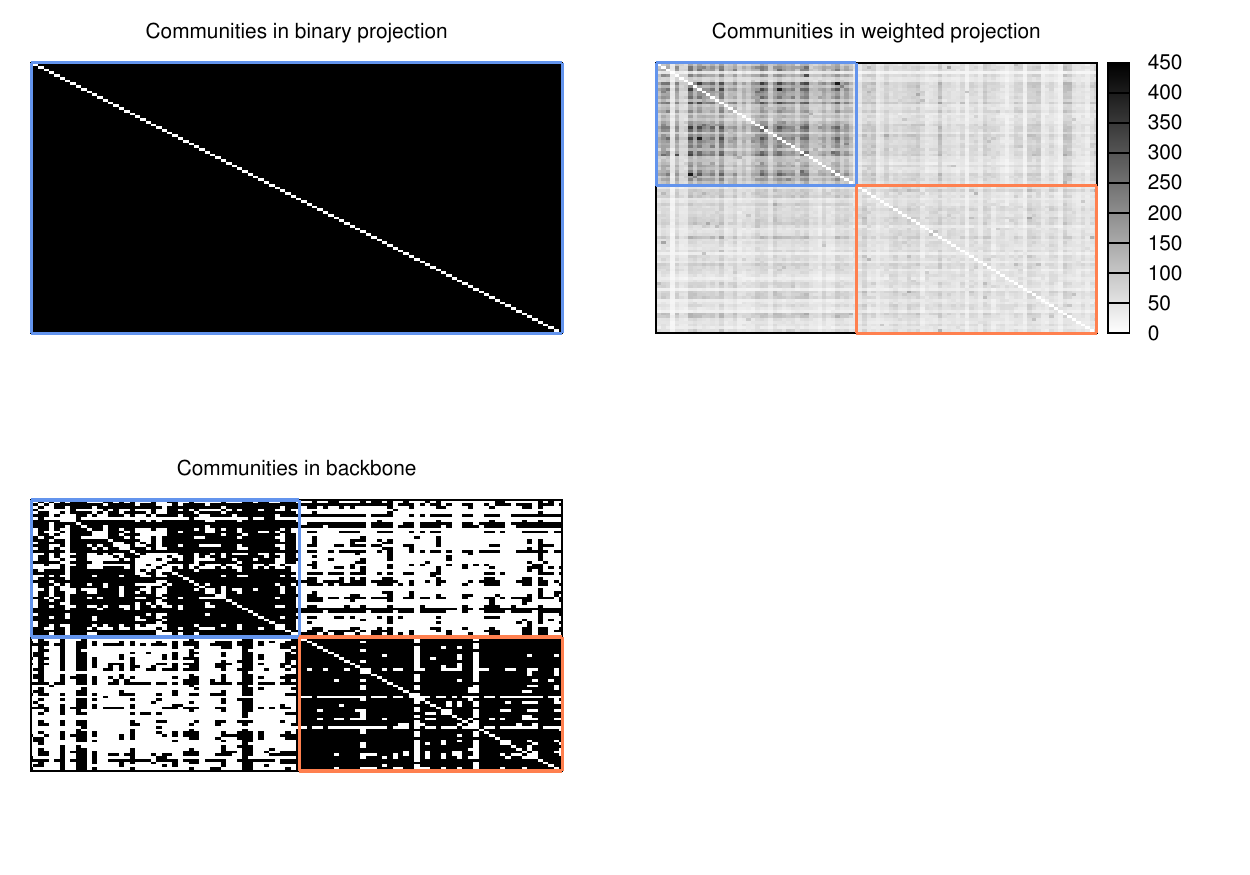}
	\caption{108\textsuperscript{th} U.S. Senate network: The adjacency matrices of the binary projection, the weighted projection and the backbone (see label of each). The identified communities are highlighted by coloured rectangles. As is evident, the backbone extraction preserves the intra community links, whereas many of the inter community connections are discarded. Senators are ordered according to their community association (first democrats then republicans) and then alphabetically by last name.}\label{im:groups108}
\end{figure}

If the backbone is extracted by means of the method introduced in the previous section, besides redundant information being discarded, a more pronounced community structure is revealed. The majority of the edges connect senators from the same party with relatively few edges connecting senators from different parties (see Fig.~\ref{im:groups108}, bottom left plot). Figure~\ref{im:groups108} shows the adjacency matrices of the binary projection (top left), the weighted projection (top right) and the backbone (bottom left) with a black square indicating the presence of an edge. The identified communities are highlighted by coloured rectangles.

The binary one-mode projection (Fig. \ref{im:groups108}, top left) was the complete graph $K_{100}$ and hence communities could not be identified. Consideration of the edge weights (Fig. \ref{im:groups108}, top right) resulted in the detection of two communities. Two communities were also detected in the backbone (Fig. \ref{im:groups108}, bottom left), however, they were more pronounced than in the weighted projection. The backbone achieved the highest modularity of 0.24 compared to 0.082 in the weighted projection. Note that the modularity ranges between 0 and 1. 94\% of the senators associated with the first community were democratic senators, whereas 96\% of the senators associated with the second community were republican. The republican senators who were associated with the first community were Lincoln Chafee, Susan M. Collins and Olympia J. Snowe, whereas the democratic senators associated with the second community were Kent Conrad and Zell Miller. 

Researching these senators revealed the following: Lincoln Chafee was a member of the Republican Party until 2007, when he became an independent, later, in 2013, joining the Democratic Party. Susan Collins, a known moderate member of the Republican Party, is considered bipartisan. Like Collins, Olympia Snowe is also known to be strongly bipartisan. Kent Conrad was found to be more conservative than most other democratic politicians, hence his association with the second community of mostly republican party members. Zell Miller was also found to be conservative. In 2004 he backed President George W. Bush over the democratic nominee.

% % % % % % % % % % % % % % % % % % % % % % % % % % % % % % % % % % % % % % % % 
\subsection{MovieLens Tags}
The MovieLens Tag Genome dataset was collected by the University of Minnesota and is available for download at http://grouplens.org/datasets/movielens/tag-genome/. This dataset contains 9,734 movies and 1,128 tags. Tags are words assigned to movies by users of the MovieLens website. Users may use as a tag any word that they feel best describes a movie. Edges connect tags to movies and record the strength of the association of a particular movie with a particular tag. Edge weights range between 0 and 1, where 1 indicates strong relevance. In our analysis we considered a subset, including the 100 most popular tags (popularity as determined by members of the GroupLens research group, who also collected the data, http://grouplens.org/). Edges were only included if the tag relevance was greater or equal to 0.5, resulting in a network of 9,734 movies and 100 tags.

\begin{figure}[h]
	\centering
	\includegraphics[width=80mm]{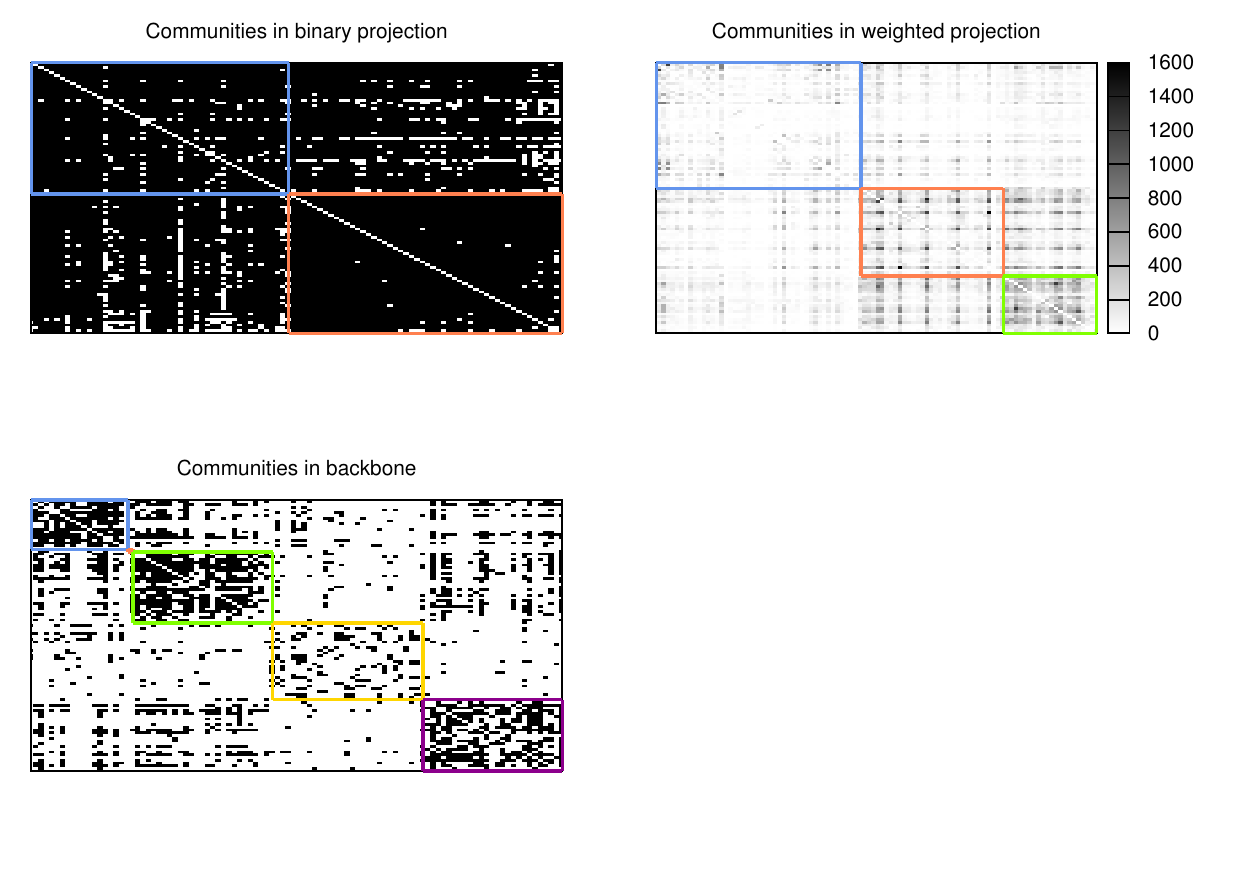}
	\caption{MovieLens Tag Genome network: The adjacency matrices of the binary projection, the weighted projection and the backbone (see label of each). The identified communities are highlighted by coloured rectangles. Again, the backbone extraction preserves the intra community connections, discarding many of the inter community connections. More information on the community membership of the different tags can be found in the supplementary movie TagNetwork.mp4.}\label{im:groupsMovieLens}
\end{figure}

Extracting \hfill the\hfill backbone\hfill of\hfill the\hfill tag-tag\hfill projection\\ should not only remove redundant information but additionally preserve connections between similar tags and remove those between unrelated tags, yielding more pronounced communities. For the same reasons as given for the U.S. Senate network, we chose a threshold of three standard deviations for the backbone extraction. Figure \ref{im:groupsMovieLens} shows the adjacency matrices of the binary projection (top left), the weighted projection (top right) and the backbone (bottom left). The community detection algorithm was able to identify two communities in the binary projection with a modularity of 0.018, three communities in the weighted projection with modularity 0.13 and five communities in the backbone with modularity 0.26. 

\begin{figure}[h]
	\centering
	\includegraphics[width=80mm]{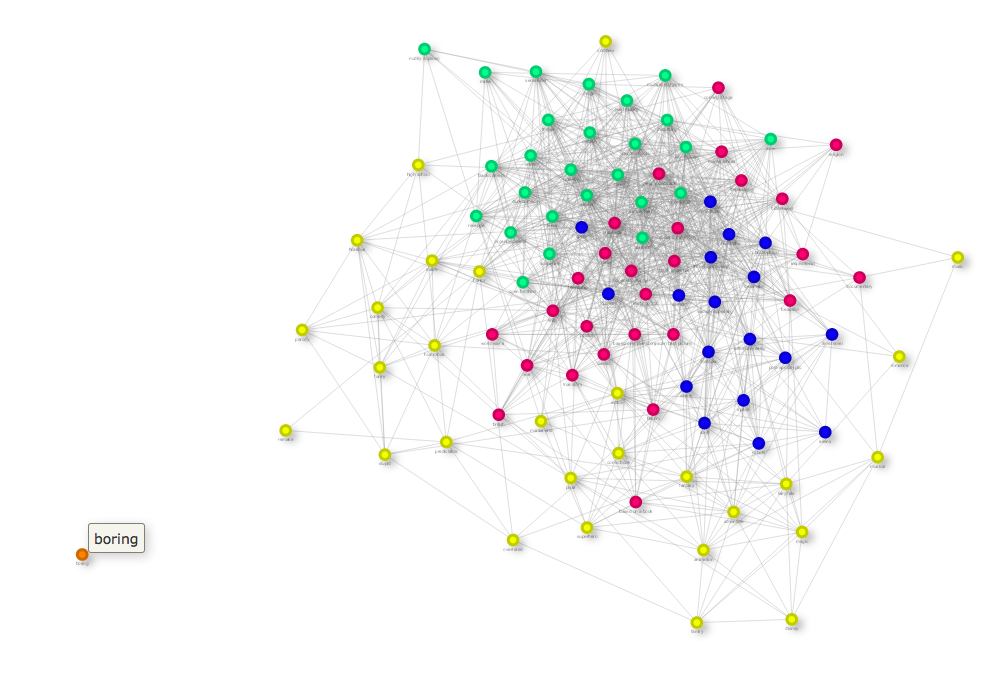}
	\caption{The backbone network of the Tag-tag projection shows an isolated node that forms a community by itself. The different colours represent community membership.}\label{im:IsolatedNode}
\end{figure}

\begin{figure}[h]
	\centering
	\includegraphics[width=80mm]{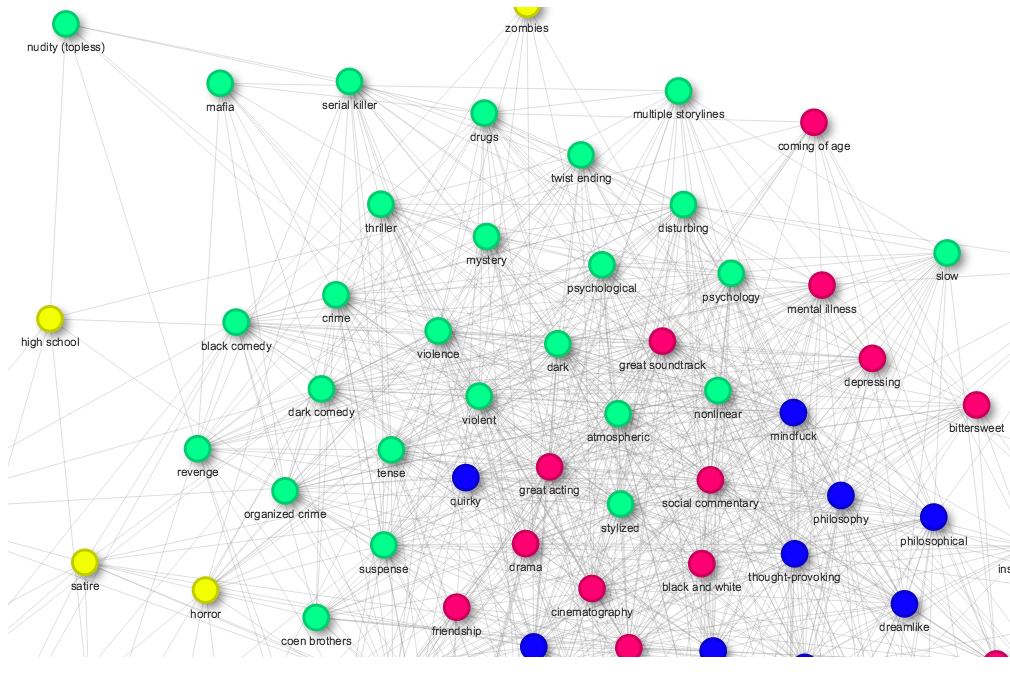}
	\caption{Looking closer at one of the communities in the backbone, we find that similar tags tend to form significant connections.}\label{im:Community}
\end{figure}

Interestingly one of the tags is isolated in the backbone, forming a community by itself (see Fig. \ref{im:IsolatedNode}). This isolated tag, labelled boring, did not form any significant connections to other nodes in the network. The other four communities each contain tags that are very similar, for instance, the tags comedy, funny, animation, satire and pixar are members of the same community (see Fig. \ref{im:Community}).

% % % % % % % % % % % % % % % % % % % % % % % % % % % % % % % % % % % % % % % % 
\section{Conclusion}\label{sec:conclusions}
Since large complex networks are often difficult to analyse, it is important to develop efficient methods that can identify redundant information to shrink the data volume. In this paper, we have shown a simple and effective way of extracting the backbone of a one-mode projection, making use of the fact that the edge weights in most projected random networks follow a Poisson binomial distribution, thus drastically reducing the computation time needed for backbone extraction.

We used two different real world networks to demonstrate the backbone as an aid to detection of communities. In the process of extracting the backbone, connections between nodes of the same community are preserved, whereas inter community links are mostly insignificant and hence discarded. Thus the backbone displays more pronounced communities leading to detection algorithms achieving higher modularity scores. 

\bibliographystyle{model1-num-names}
\bibliography{BackboneExtraction}

\begin{thebibliography}{25}
\expandafter\ifx\csname natexlab\endcsname\relax\def\natexlab#1{#1}\fi
\providecommand{\url}[1]{\texttt{#1}}
\providecommand{\href}[2]{#2}
\providecommand{\path}[1]{#1}
\providecommand{\DOIprefix}{doi:}
\providecommand{\ArXivprefix}{arXiv:}
\providecommand{\URLprefix}{URL: }
\providecommand{\Pubmedprefix}{pmid:}
\providecommand{\doi}[1]{\href{http://dx.doi.org/#1}{\path{#1}}}
\providecommand{\Pubmed}[1]{\href{pmid:#1}{\path{#1}}}
\providecommand{\bibinfo}[2]{#2}
\ifx\xfnm\relax \def\xfnm[#1]{\unskip,\space#1}\fi
%Type = Article
\bibitem[{Glattfelder and Battiston(2009)}]{Glattfelder2009}
\bibinfo{author}{J.~B. Glattfelder}, \bibinfo{author}{S.~Battiston},
\newblock \bibinfo{title}{{Backbone of complex networks of corporations: The
  flow of control}},
\newblock \bibinfo{journal}{Physical Review E} \bibinfo{volume}{80}
  (\bibinfo{year}{2009}) \bibinfo{pages}{1--12}.
%Type = Article
\bibitem[{Slater(2009)}]{Slater2009}
\bibinfo{author}{P.~B. Slater},
\newblock \bibinfo{title}{{A two-stage algorithm for extracting the multiscale
  backbone of complex weighted networks}},
\newblock \bibinfo{journal}{Proceedings of the National Academy of Sciences}
  \bibinfo{volume}{106} (\bibinfo{year}{2009}) \bibinfo{pages}{E66}.
%Type = Article
\bibitem[{Scutari and Nagarajan(2013)}]{Scutari2013}
\bibinfo{author}{M.~Scutari}, \bibinfo{author}{R.~Nagarajan},
\newblock \bibinfo{title}{{Identifying significant edges in graphical models of
  molecular networks}},
\newblock \bibinfo{journal}{Artificial Intelligence in Medicine}
  \bibinfo{volume}{57} (\bibinfo{year}{2013}) \bibinfo{pages}{207--217}.
%Type = Article
\bibitem[{Zhang et~al.(2014)Zhang, Zhang, Zhao, Wang, and Zhu}]{Zhang2014}
\bibinfo{author}{X.~Zhang}, \bibinfo{author}{Z.~Zhang},
  \bibinfo{author}{H.~Zhao}, \bibinfo{author}{Q.~Wang},
  \bibinfo{author}{J.~Zhu},
\newblock \bibinfo{title}{{Extracting the globally and locally adaptive
  backbone of complex networks.}},
\newblock \bibinfo{journal}{PLoS One} \bibinfo{volume}{9}
  (\bibinfo{year}{2014}) \bibinfo{pages}{e100428}.
%Type = Article
\bibitem[{Serrano et~al.(2009)Serrano, Bogu{\~{n}}{\'{a}}, and
  Vespignani}]{Serrano2009}
\bibinfo{author}{M.~A. Serrano}, \bibinfo{author}{M.~Bogu{\~{n}}{\'{a}}},
  \bibinfo{author}{A.~Vespignani},
\newblock \bibinfo{title}{{Extracting the multiscale backbone of complex
  weighted networks.}},
\newblock \bibinfo{journal}{Proceedings of the National Academy of Sciences of
  the United States of America} \bibinfo{volume}{106} (\bibinfo{year}{2009})
  \bibinfo{pages}{6483--6488}.
%Type = Article
\bibitem[{Foti et~al.(2011)Foti, Hughes, and Rockmore}]{Foti2011}
\bibinfo{author}{N.~J. Foti}, \bibinfo{author}{J.~M. Hughes},
  \bibinfo{author}{D.~N. Rockmore},
\newblock \bibinfo{title}{{Nonparametric sparsification of complex multiscale
  networks}},
\newblock \bibinfo{journal}{PLoS One} \bibinfo{volume}{6}
  (\bibinfo{year}{2011}).
%Type = Article
\bibitem[{Zhang et~al.(2013)Zhang, Zeng, and Shang}]{Zhang2013b}
\bibinfo{author}{Q.-M. Zhang}, \bibinfo{author}{A.~Zeng},
  \bibinfo{author}{M.-S. Shang},
\newblock \bibinfo{title}{{Extracting the information backbone in online
  system.}},
\newblock \bibinfo{journal}{PLoS One} \bibinfo{volume}{8}
  (\bibinfo{year}{2013}) \bibinfo{pages}{e62624}.
%Type = Article
\bibitem[{Neal(2013)}]{Neal2013}
\bibinfo{author}{Z.~Neal},
\newblock \bibinfo{title}{{Identifying statistically significant edges in
  one-mode projections}},
\newblock \bibinfo{journal}{Social Network Analysis and Mining}
  \bibinfo{volume}{3} (\bibinfo{year}{2013}) \bibinfo{pages}{915--924}.
%Type = Article
\bibitem[{Neal(2014)}]{Neal2014}
\bibinfo{author}{Z.~Neal},
\newblock \bibinfo{title}{{The backbone of bipartite projections: Inferring
  relationships from co-authorship, co-sponsorship, co-attendance and other
  co-behaviors}},
\newblock \bibinfo{journal}{Social Networks} \bibinfo{volume}{39}
  (\bibinfo{year}{2014}) \bibinfo{pages}{84--97}.
%Type = Article
\bibitem[{Borgatti and Everett(1997)}]{Borgatti1997}
\bibinfo{author}{S.~P. Borgatti}, \bibinfo{author}{M.~G. Everett},
\newblock \bibinfo{title}{{Network analysis of 2-mode data}},
\newblock \bibinfo{journal}{Social Networks} \bibinfo{volume}{19}
  (\bibinfo{year}{1997}) \bibinfo{pages}{243--269}.
%Type = Inproceedings
\bibitem[{Liebig and Rao(2014)}]{Liebig2014}
\bibinfo{author}{J.~Liebig}, \bibinfo{author}{A.~Rao},
\newblock \bibinfo{title}{{Identifying influential nodes in bipartite networks
  using the clustering coefficient}},
\newblock in: \bibinfo{booktitle}{Proceedings of the 2014 Tenth International
  Conference on Signal-Image Technology and Internet-Based Systems},
  \bibinfo{year}{2014}, pp. \bibinfo{pages}{323--330}.
%Type = Article
\bibitem[{Liebig and Rao(2016)}]{Liebig2016}
\bibinfo{author}{J.~Liebig}, \bibinfo{author}{A.~Rao},
\newblock \bibinfo{title}{{Predicting item popularity: Analysing local
  clustering behaviour of users}},
\newblock \bibinfo{journal}{Physica A: Statistical Mechanics and its
  Applications} \bibinfo{volume}{442} (\bibinfo{year}{2016})
  \bibinfo{pages}{523--531}.
%Type = Article
\bibitem[{Newman(2001)}]{Newman2001a}
\bibinfo{author}{M.~E.~J. Newman},
\newblock \bibinfo{title}{{The structure of scientific collaboration
  networks.}},
\newblock \bibinfo{journal}{Proceedings of the National Academy of Sciences of
  the United States of America} \bibinfo{volume}{98} (\bibinfo{year}{2001})
  \bibinfo{pages}{404--409}.
%Type = Article
\bibitem[{Latapy et~al.(2008)Latapy, Magnien, and Vecchio}]{Latapy2008}
\bibinfo{author}{M.~Latapy}, \bibinfo{author}{C.~Magnien},
  \bibinfo{author}{N.~D. Vecchio},
\newblock \bibinfo{title}{{Basic notions for the analysis of large two-mode
  networks}},
\newblock \bibinfo{journal}{Social Networks} \bibinfo{volume}{30}
  (\bibinfo{year}{2008}) \bibinfo{pages}{31--48}.
%Type = Book
\bibitem[{Forbes et~al.(2011)Forbes, Evans, Hastings, and Peacock}]{Forbes2011}
\bibinfo{author}{C.~Forbes}, \bibinfo{author}{M.~Evans},
  \bibinfo{author}{N.~Hastings}, \bibinfo{author}{B.~Peacock},
  \bibinfo{title}{{Statistical distributions}}, \bibinfo{publisher}{{John Wiley
  \& Sons, Inc.}}, \bibinfo{address}{Hoboken}, \bibinfo{year}{2011}.
%Type = Article
\bibitem[{Wang(1993)}]{Wang1993}
\bibinfo{author}{Y.~H. Wang},
\newblock \bibinfo{title}{On the numer of successes in independent trials},
\newblock \bibinfo{journal}{{Statistica Sinica}} \bibinfo{volume}{2}
  (\bibinfo{year}{1993}) \bibinfo{pages}{295--312}.
%Type = Book
\bibitem[{Lando(2003)}]{Lando2003}
\bibinfo{author}{S.~K. Lando}, \bibinfo{title}{{Lectures on generating
  functions}}, \bibinfo{publisher}{{American Mathematical Society}},
  \bibinfo{address}{Providence}, \bibinfo{year}{2003}.
%Type = Book
\bibitem[{Newman(2010)}]{Newman2010}
\bibinfo{author}{M.~E.~J. Newman}, \bibinfo{title}{{Networks: An
  introduction}}, \bibinfo{edition}{1} ed., \bibinfo{publisher}{Oxford
  University Press}, \bibinfo{address}{Oxford}, \bibinfo{year}{2010}.
%Type = Article
\bibitem[{{Le Cam}(1960)}]{LeCam1960}
\bibinfo{author}{L.~{Le Cam}},
\newblock \bibinfo{title}{{An approximation theorem for the Poisson binomial
  distribution}},
\newblock \bibinfo{journal}{Pacific Journal of Mathematics}
  \bibinfo{volume}{10} (\bibinfo{year}{1960}) \bibinfo{pages}{1181--1197}.
%Type = Misc
\bibitem[{{KONECT}(2014)}]{Konect2014}
\bibinfo{author}{{KONECT}}, \bibinfo{title}{{Networks}}, \bibinfo{year}{2014}.
  \URLprefix \url{http://konect.uni-koblenz.de/networks/}.
%Type = Article
\bibitem[{Fowler(2006{\natexlab{a}})}]{Fowler2006}
\bibinfo{author}{J.~H. Fowler},
\newblock \bibinfo{title}{{Connecting the congress: A study of cosponsorship
  networks}},
\newblock \bibinfo{journal}{Political Analysis} \bibinfo{volume}{14}
  (\bibinfo{year}{2006}{\natexlab{a}}) \bibinfo{pages}{456--487}.
%Type = Article
\bibitem[{Fowler(2006{\natexlab{b}})}]{Fowler2006a}
\bibinfo{author}{J.~H. Fowler},
\newblock \bibinfo{title}{{Legislative cosponsorship networks in the US House
  and Senate}},
\newblock \bibinfo{journal}{Social Networks} \bibinfo{volume}{28}
  (\bibinfo{year}{2006}{\natexlab{b}}) \bibinfo{pages}{454--465}.
%Type = Article
\bibitem[{Vig et~al.(2012)Vig, Sen, and Riedl}]{Vig2012}
\bibinfo{author}{J.~Vig}, \bibinfo{author}{S.~Sen}, \bibinfo{author}{J.~Riedl},
\newblock \bibinfo{title}{{The tag genome: Encoding community knowledge to
  support novel interaction}},
\newblock \bibinfo{journal}{ACM Transactions on Interactive Intelligent Systems
  (TiiS)} \bibinfo{volume}{2} (\bibinfo{year}{2012}) \bibinfo{pages}{13}.
%Type = Article
\bibitem[{Newman(2006)}]{Newman2006}
\bibinfo{author}{M.~E.~J. Newman},
\newblock \bibinfo{title}{{Finding community structure in networks using the
  eigenvectors of matrices}},
\newblock \bibinfo{journal}{Physical Review E} \bibinfo{volume}{74}
  (\bibinfo{year}{2006}) \bibinfo{pages}{036104}.
%Type = Article
\bibitem[{Cs\'{a}rdi and Nepusz(2006)}]{Csardi2006}
\bibinfo{author}{G.~Cs\'{a}rdi}, \bibinfo{author}{T.~Nepusz},
\newblock \bibinfo{title}{{The igraph software package for complex network
  research.}},
\newblock \bibinfo{journal}{InterJournal, Complex Systems}
  \bibinfo{volume}{1695} (\bibinfo{year}{2006}) \bibinfo{pages}{1--9}.

\end{thebibliography}
	
\end{document}